\documentclass[twocolumn,showpacs,preprintnumbers,amsmath,amssymb,superscriptaddress]{revtex4}
\usepackage{graphicx}
\usepackage{dcolumn}
\usepackage{bm}

\begin{document}

\title{Combined chips for atom-optics}

\author{A. G\"unther}
\homepage{http://www.pit.physik.uni-tuebingen.de/zimmermann/}
\email{aguenth@pit.physik.uni-tuebingen.de}
\author{M. Kemmler}
\author{S. Kraft}
\affiliation{Physikalisches Institut der Universit\"at T\"ubingen,
Auf der Morgenstelle 14, 72076 T\"ubingen, Germany}
\author{C. J. Vale}
\affiliation{School of Physical Sciences, University of
Queensland, Brisbane, Queensland 4072, Australia}
\author{C. Zimmermann}
\author{J. Fort\'agh}
\affiliation{Physikalisches Institut der Universit\"at T\"ubingen,
Auf der Morgenstelle 14, 72076 T\"ubingen, Germany}

\date{\today}

\begin{abstract}
We present experiments with Bose-Einstein condensates on a
combined atom chip.  The combined structure consists of a
large-scale ``carrier chip'' and smaller ``atom-optics chips'',
containing micron-sized elements.  This allows us to work with
condensates very close to chip surfaces without suffering from
fragmentation or losses due to thermally driven spin flips.
Precise three-dimensional positioning and transport with constant
trap frequencies are described.  Bose-Einstein condensates were
manipulated with submicron accuracy above atom-optics chips. As an 
application of atom chips, a direction sensitive magnetic
field microscope is demonstrated.
\end{abstract}

\maketitle

\section{Introduction}
\label{sec:level1}

Interest in phase coherent experiments with ultracold quantum
gases has motivated numerous theoretical and experimental
investigations into microscopic magnetic traps
\cite{Weinstein1995a}. An intriguing feature of microtraps is the
possibility of constructing complex potentials using the magnetic
field of micro-structured conductors on a chip, as well as the
precise spatial and temporal control of these potentials. Specific
atom-optical experiments have been discussed, such as
Bose-Josephson junctions in double well potentials
\cite{Smerzi1997a}, tunneling through a quantum dot in magnetic
waveguides \cite{Paul2005a} or even scenarios for creating
entangled pairs of atoms \cite{Dorner2003a}. A central goal is the
preparation of matter waves in microtraps with potential
structures on the micron scale, comparable to the healing length
of the condensate, which will allow for tunneling on
experimentally viable timescales. Current experiments can readily
produce Bose-Einstein condensates in magnetic microtraps
\cite{Ott2001a,Haensel2001a}, however, the potentials used to date
have been too large scaled to reach the tunneling regime. To go
beyond this, condensates must be moved closer to field generating
elements, that is, to within a few microns of the chip surface.

With atom/conductor separations of a few microns, Johnson-noise
induced spin-flip losses reduce the life time of the atomic cloud
\cite{Henkel1999a,Jones2004a}. Furthermore, at distances extending
to several tens of microns, irregularities of the trapping
potential, arising from geometrical imperfections of the conductor
\cite{Estve2004a} and the corresponding spatial fluctuation of
magnetic field \cite{Kraft2002a}, cause fragmentation of atomic
clouds.

In this paper we present a way to overcome these limiting
surface effects which allows one to work at micron distances from
the surface. A smooth waveguide potential is produced by a large
scaled ``carrier chip'' situated several hundreds of microns from
the surface of further ``atom optics chips''. At this distance,
surface effects from the carrier are negligible \cite{Ott2003a}.
The carrier chip is used for micro-positioning a condensate.
Attached to the surface of the carrier, additional atom-optics
chips based on micron scaled conductor patterns are used for
steep and thin potential barriers. Due to the small amount of metal, 
Johnson-noise induced spin-flip rates are minimized even at close 
distances to the chip surface \cite{Henkel2003}. Since atoms are confined in a
smooth waveguide and interact only with short sections of cross
conductors at micron distances, stable conditions for coherent
atom optics are achieved.

\section{\label{sec:level5}Three-dimensional confinement on a chip}

When manipulating condensates with micron-sized electromagnets,
the exact position of the atoms with respect to these elements
becomes crucial. Previous micro traps have used large external
electromagnets to generate the so called bias field that is
essential for the microtrap concept \cite{Folman2002a}. Because
of the uncertainties in their geometry, accurate calculation of
the trap parameters and position becomes difficult. In order to
overcome this, we have developed a chip which includes all field
generating elements necessary for three-dimensional confinement
and manipulation. The micro fabricated wire pattern defines a
precise conductor geometry and the trap parameters and position
can be computed with high accuracy. The conductor configuration
was designed to provide three-dimensional positioning and
transport of condensates, while keeping the trap frequencies
constant.

\subsection{\label{sec:level5}Radial confinement and positioning}

\begin{figure}
\centerline{\scalebox{0.9}{\includegraphics{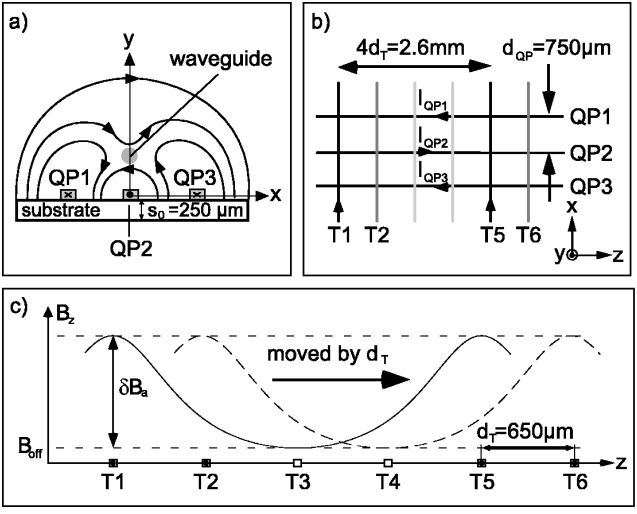}}}
\caption{a) Three wire configuration on a chip used for a magnetic
waveguide potential. The central conductor QP2 carries current
opposite to the outer conductors QP1 and QP3. A waveguide is
formed above the chip, as indicated by the magnetic field lines.
Marked also is the thickness of the substrate $s_0=250\mu\text{m}$
as used for the carrier chip in our experiment. b) Conductor
geometry for a microtrap with three dimensional confinement. A
waveguide potential is generated by the conductors QP1, QP2, and
QP3. The magnetic field of a pair of perpendicular conductors T1
and T5 provide axial confinement with non-vanishing offset field.
An equidistant set of perpendicular conductors ("transport
conductors") can be used for an axial displacement. The geometry
of conductors is marked: the distance between the QP wires
$d_\text{QP}$ is $750\mu\text{m}$ and the distance between the
transport wires is $d_T=650\mu\text{m}$. c) Translation of the
trap. If the current in T1 and T5 is replaced by a current in T2 and
T6, the trap is axially displaced by the distance of
$d_T=650\mu\text{m}$.} \label{fig:Mikrofallenprinzip}
\end{figure}

A magnetic waveguide potential can be realized by three parallel
conductors as shown in Figure \ref{fig:Mikrofallenprinzip}a and \ref{fig:Mikrofallenprinzip}b. The
magnetic field of the center conductor QP2, driven with a current
of $I_\text{QP2}$, is superimposed with the bias field
$\bm{B}_{\bot}$, generated by two other parallel conductors QP1
and QP3 (``quadrupole wires'').

The inhomogeneous bias field is given by
\begin{eqnarray}
\begin{aligned}
{\bm{B}_{\bot}}&=\frac{\mu_{0}}{2\pi}\left( \frac{I_{\text{QP1}}y}{\left(d_{\text{QP}}+x\right)^2 + y^2} + \frac{I_{\text{QP3}}y}{\left(d_{\text{QP}}-x\right)^2 + y^2} \right)\bm{e}_{x}\\
&-\frac{\mu_{0}}{2\pi}\left(\frac{I_{\text{QP1}}\left(d_{\text{QP}}+x\right)}{\left(d_{\text{QP}}+x\right)^2 + y^2} - \frac{I_{\text{QP3}}\left(d_{\text{QP}}-x\right)}{\left(d_{\text{QP}}-x\right)^2 + y^2} \right)\bm{e}_{y}\;.
\end{aligned}
\end{eqnarray}
$d_{QP}$ is the distance between the conductors QP1 and QP2, (and
QP2 and QP3) and $I_{QPi}$, $i = 1,2,3$ are the respective
currents. The waveguide forms in a line, where the bias field and
the field of the central conductor cancel each other. This
requires the current in the outer conductors to be opposite in
direction to the current in the central conductor. We follow the convention that $I_{\text{QP2}}$ is positive for currents in +z direction, and $I_{\text{QP1/QP3}}$ is positive in -z direction.

Applying the same current in QP1 and QP3, the $y$-component of the
bias field vanishes along the $y$-axis, and the resulting bias field is
\begin{equation}
{\bm{B}_{\bot}}\left(x=0,y\right)=\frac{\mu_{0}}{\pi}\frac{yI_{\text{QP1}}}{d_{\text{QP}}^2+y^2}\bm{e}_{x}\;.
\end{equation}
The waveguide forms at a distance of
\begin{equation}
y_{0}=d_{\text{QP}}\sqrt{\frac{1}{2I_{\text{QP1}}/I_{\text{QP2}}-1}}
\label{eq:Waveguide_Distance}
\end{equation}
on the $y$-axis (Figure \ref{fig:Mikrofallenprinzip}a). A trap is
only formed if $|I_{\text{QP1}}|\geq |I_{\text{QP2}}/2|$. The
radial field gradient $a_{r}$ in the trap center of the
three-wire configuration is given by
\begin{eqnarray}
\begin{aligned}
\label{eq:gradient}
a_{r} &= \frac{\mu_{0}}{2\pi}\frac{I_{\text{QP2}}}{y_0^2}\frac{2}{1+\left(y_0/d_{\text{QP}}\right)^2}\;.
\end{aligned}
\end{eqnarray}

\begin{figure}
\centerline{\scalebox{0.95}{\includegraphics{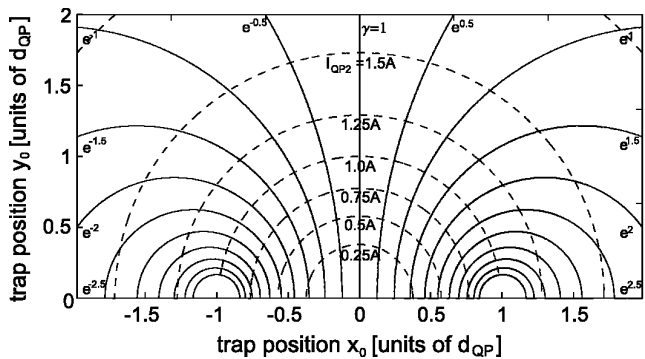}}}
\caption{Position of the waveguide in the $x,y$-plane. Solid lines
are trajectories for a constant ratio of
$\gamma=I_\text{QP1}/I_\text{QP3}$ and increasing $I_\text{QP2}$.
Dashed lines are trajectories for constant $I_\text{QP2}$ and
varying $\gamma=I_\text{QP1}/I_\text{QP3}$. The sum of the currents
$I_\text{QP1}$+$I_\text{QP3}$=2A was kept constant in this plot. }
\label{fig:Trapposition}
\end{figure}

In the general case, the $x,y$-position of the waveguide is determined by the currents
$I_{\text{QP1}}$, $I_{\text{QP2}}$, and $I_{\text{QP3}}$:
\begin{eqnarray}
\begin{aligned}
x_0 &= \frac{d_\text{QP}}{2}\frac{I_\text{QP1}-I_\text{QP3}}{I_\text{QP1}+I_\text{QP2}+I_\text{QP3}}\\
y_0 &= \frac{d_\text{QP}}{2}\frac{\left(4I_\text{QP1}I_\text{QP3}-(I_\text{QP1}+2I_\text{QP2}+I_\text{QP3})^2\right)^{1/2}}{I_\text{QP1}+I_\text{QP2}+I_\text{QP3}}\;. \end{aligned}
\end{eqnarray}
The trap can be positioned within a large area above the chip by
changing the currents $I_{\text{QP1}}$, $I_{\text{QP3}}$, and
$I_{\text{QP2}}$. Figure \ref{fig:Trapposition} shows trajectories
corresponding to a constant ratio of $I_{\text{QP1}}$ and
$I_{\text{QP3}}$ (solid lines) and constant $I_{\text{QP2}}$
(dashed lines) while keeping the sum
$I_{\text{QP1}}$+$I_{\text{QP3}}$ at a constant current of 2A. The
values are realistic for experiments discussed later.

\subsection{\label{sec:level5} Axial confinement and translation of the trap center}

The waveguide produced by the three conductors will suffer from
Majorana spin flips \cite{Sukumar1997a} because the magnetic field
vanishes in the center of the waveguide. However, a magnetic
offset field $\bm{B}_{\text{off}}$ along the symmetry axis,
stabilizes the trap against these losses. The nonzero absolute
value of the magnetic field $\bm{B}_{\text{off}}$ in the trap
center also changes the radial potential shape from linear to
parabolic \cite{Pritchard1983a}. The radial confinement is then
characterized by the oscillation frequency \mbox{$\omega_r=\sqrt{g_F
\mu_B m_F/m B_{\text{off}}}\cdot a_r$}, with the Land\'e factor $g_F$, the Bohr magneton $\mu_B$, the magnetic quantum number $m_F$, and m the mass of the atom. To allow for
three dimensional confinement, an inhomogeneous offset field is
generated by two conductors, T1 and T5 in Figure
\ref{fig:Mikrofallenprinzip}b, perpendicular to the waveguide ("transport wires"). The
distance between T1 and T5 is 4$d_T$. For identical currents
$I_\text{T}$ in T1 and T5, the magnetic field is described by its
components

\begin{eqnarray}
\label{eq:FieldOfTransport}
\textstyle{B_x}&\textstyle{=}&\textstyle{0}\\
\textstyle{B_y}&\textstyle{=}&\textstyle{\frac{\mu_0}{2\pi}I_T\left(\frac{2d_T+z}{(s_0+y_0)^2+\left(2d_T+z\right)^2}-\frac{2d_T-z}{(s_0+y_0)^2+\left(2d_T-z\right)^2}\right)}\nonumber\\
\textstyle{B_z}&\textstyle{=}&\textstyle{-\frac{\mu_0}{2\pi}I_T\left(\frac{s_0+y_0}{(s_0+y_0)^2+\left(2d_T+z\right)^2}+\frac{s_0+y_0}{(s_0+y_0)^2+\left(2d_T-z\right)^2}\right)}\nonumber\;.
\end{eqnarray}
In our setup the transport wires are located at 
the back side of the substrate separated from the QP wires by $s_0$. 
The $B_z$ component provides the overall non-vanishing
offset field for the waveguide. Near the minimum of $B_z$, between
T1 and T5, the potential is approximately parabolic, with an axial
oscillation frequency $\omega_a$ (Eq. \ref{eq:trapfrequencies}) and an offset field value
\begin{equation}
B_{\text{off}}=\frac{\mu_0}{\pi}\frac{I_T(s_0+y_0)}{(s_0+y_0)^2+4d_T^2}\;.
\end{equation}
The axial potential is plotted in Figure
\ref{fig:Mikrofallenprinzip}c. The field component $B_y$ causes a
small rotation of the waveguide about the $y$-axis. The rotation
angle at the bottom of the trap is estimated by
\begin{equation}
\alpha\approx-\frac{\mu_0}{\pi}\frac{I_T}{a_r}\frac{y_0^2-4d_T^2}{\left(y_0^2+4d_T^2\right)^2}\;.
\end{equation}
The trapping potential generated by the five conductors QP1, QP2,
QP3, T1, and T5 is characterized by the field curvatures $b_i$
\begin{eqnarray}
\begin{aligned}
b_{\text{axial}}&=\frac{\mu_0}{2\pi}I_T (s_0+y_0) \frac{48d_T^2-4(s_0+y_0)^2}{\left((s_0+y_0)^2+4d_T^2\right)^3}\\
b_{\text{radial}}&=\frac{a_r^2}{B_{\text{off}}}=\frac{\mu_0}{2\pi}\frac{2I_{\text{QP2}}^2}{y_0^4 I_T (s_0+y_0)}\frac{(s_0+y_0)^2+4d_T^2}{(1+\left(r_0/d_{\text{QP}}\right)^2)^2}\;,
\end{aligned}
\end{eqnarray}
and the resulting trap frequencies $\omega_i$, $i = $axial,radial
\begin{equation}
\label{eq:trapfrequencies}
\omega_i=\sqrt{\frac{g_F m_F \mu_B}{m}b_i}\;.
\end{equation}
The above expressions assume infinite conductor lengths. The
finite length of the conductors on the real chip (Figure \ref{fig:Atomchip})
results in a deviation of the axial and radial trap frequencies by
5 and 4 percent, respectively.

The five-wire configuration achieves full, three dimensional
confinement of atoms using magnetic fields generated by the chip
only. Precise control of the position ($x_0,y_0$) and parameters
of the trap is therefore possible. The $z$-position of the trap is
controlled by further transport wires T1 - T8.  A trap generated by T1 and T5 will have identical
parameters to one generated by T2 and T6 (Figure
\ref{fig:Mikrofallenprinzip}). One trap can be continuously
transformed into the other by reducing the current in one of the
pairs while increasing it in the other. Thereby, the trap center
shifts by $d_T$, the distance between two neighboring conductors
$T_\text{i}$ and $T_\text{i+1}$. While in general the axial trap
frequencies change during the transport, we have calculated an
optimum trajectory of the currents $I_{\text{T2}}(I_{\text{T1}})$,
the same as for $I_{\text{T6}}(I_{\text{T5}})$, so that the axial
trapping frequency remains constant.

\begin{figure}
\centerline{\scalebox{0.87}{\includegraphics{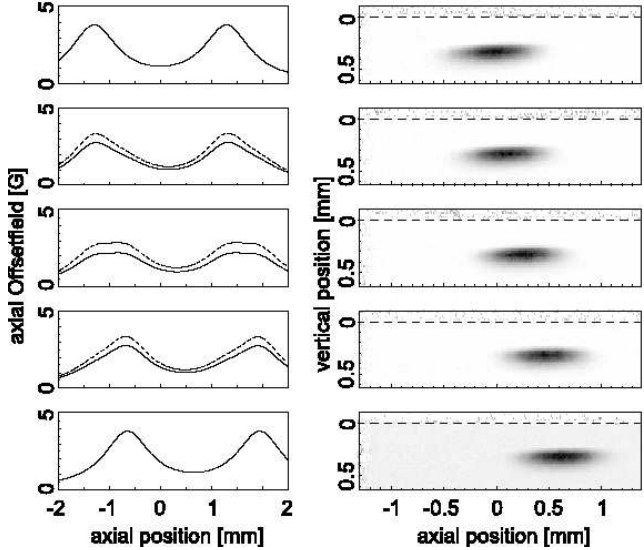}}}
\caption{(Left) The axial magnetic potential for five
displacements ($0, 162.5\mu\text{m}, 325\mu\text{m},
487,5\mu\text{m}, 650\mu\text{m}$) during transport over
$d_T=650\mu\text{m}$ (solid line). Matching the currents in T1,
T2, T4 and T5 (Eq. \ref{eq:CurrentForSmootTransport}) results in a smooth transport without
changing the trap frequencies. (Right) Absorption images of
thermal clouds for the corresponding displacements. The dashed
line shows the chip surface. In this experiment the currents were
driven linearly, resulting in a variation of the axial potential
(dashed potential curves on the left). } \label{fig:Potential}
\end{figure}

On our chip the transport wires $T_\text{i}$ are spaced by
$d_T=650\mu\text{m}$ (Figure \ref{fig:Atomchip}). For a trap to be
transported at constant height of about $300\mu\text{m}$, the
optimum currents, fitted to third order, are
\begin{equation}
\label{eq:CurrentForSmootTransport}
\frac{I_\text{T2}}{I_\text{T}}=1-2.39\frac{I_\text{T1}}{I_\text{T}}+2.31\left(\frac{I_\text{T1}}{I_\text{T}}\right)^2-0.92\left(\frac{I_\text{T1}}{I_\text{T}}\right)^3\;.
\end{equation}
The axial displacement of the trap in terms of the currents is given by
\begin{equation}
z_0=d_T\left(1-1.49\frac{I_\text{T1}}{I_\text{T}}+0.49\left(\frac{I_\text{T1}}{I_\text{T}}\right)^2\right)\;.
\end{equation}
$I_{\text{T}}$ is the initial current in T1 and T5. Keeping the
axial trap frequency and the height above the chip constant by
changing the currents in the transport wires as described above,
the offset field and correspondingly the radial trap frequency
would vary by a factor of 1.14 and 0.94, respectively. The
variation of the radial trap frequency, however, can be
compensated by varying the radial gradient (Eq. \ref{eq:gradient})
of the trap through matching the currents in the three quadrupole
wires $I_\text{QP1}=I_\text{QP3}$ and $I_\text{QP2}$ while holding
the ratio $I_\text{QP1}/I_\text{QP2}$ constant. The latter
condition is necessary for the constant height (Eq. \ref{eq:Waveguide_Distance}). On the other
hand, the transport wires produce a non-vanishing $y$-component of
the magnetic field at the trap center
(Eq. \ref{eq:FieldOfTransport}), altering the $x$-position of the
trap during the transport by $\pm 1.3\mu\text{m}$. This is in
general a negligible effect, but can also be compensated by
matching the currents $I_\text{QP1}$ and $I_\text{QP3}$ while
keeping their sum constant. As a result, we are able to move the
trap in the axial direction while keeping the radial and axial
trap frequencies as well as the radial ($x,y$) position of the
trap constant. The axial trapping potential is plotted in Figure
\ref{fig:Potential}a during the optimized transport over a distance of
 $d_T$ (solid line). The curvature of the potential near the minimum does
not change, while a small change in the offset field is visible.
During transport, the axial trap depth reduces to almost half of
its initial and final values. Nonetheless, this does not set any
relevant restrictions on experiments which are operated with
sufficiently cold ensembles.

On our atom chip, we have integrated a set of eight conductors
T1-T8 which repeat periodically over the length of the chip (see
next section). Transport of atomic clouds is possible over a
distance of 1.75 cm without any change of the trap frequencies or
the radial position. The conveyor on our chip has been
developed with respect to this scaleability, allowing coherent
transport of condensates over a large distance. A Bose-Einstein
condensate of interacting particles is a nonlinear medium
sensitive even to non-resonant excitations \cite{Ott2003a}, e.g.
through changes of the trap frequencies. Coherent, excitationless
transport of the condensate thus requires constant trap
frequencies as provided by the conveyor belt described in this
article.

\section{Combined Chips}
\label{sec:level2}

In subsequent experiments, we have used a combined atom chip. This
consists of a carrier chip, which operation principle has been outlined in the
previous section, and several atom-optics chips.

\subsection{Carrier Chip}
\label{sec:level3}

\begin{figure}
\centerline{\scalebox{0.95}{\includegraphics{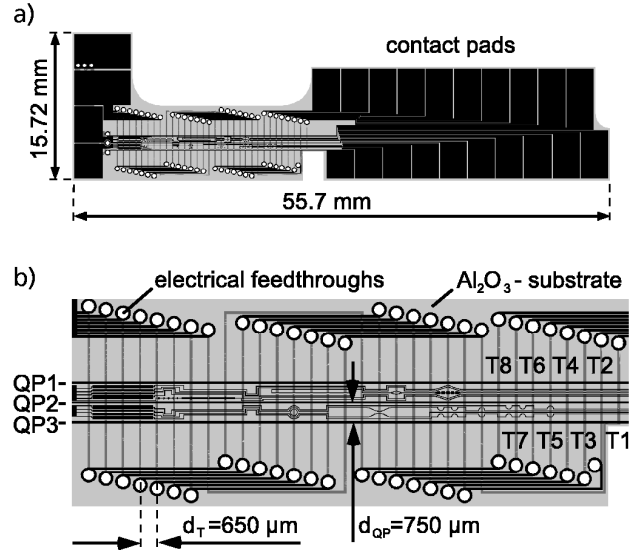}}}
\caption{Carrier chip. a) Schematic plot of the dual-layer atom
chip with conductors at both surfaces of the substrate and contact
pads on the top. b) Central part of the carrier chip. Indicated
are the conductors QP1, QP2 and QP3 on the top and the transport
wires T1-T8 on the back surface of the chip. The eight transport
wires are periodically repeated underneath the chip. Electrical
connection between the separate blocks of T1-T8 is achieved by
running the wires through laser cut holes from underneath to the
top surface of the chip and back. The additional wire pattern
between QP1 and QP2, as well as between QP2 and QP3 are not used
for the experiments described in this article. These can be used
for further microtraps as described elsewhere
\cite{Guenther2005a}.} \label{fig:Atomchip}
\end{figure}

The carrier chip is a dual-layer atom chip which produces the
entire magnetic field for three-dimensional magnetic trapping and
manipulation of Bose-Einstein condensates (Figure
\ref{fig:Atomchip}). It supports the QP and T wires described in
the previous section. Additional wire patterns shown in Figure
\ref{fig:Atomchip} between QP1 and QP2, and between QP2 and QP3
were deposited for further microtraps, however, these are not used
in the present work.

The chip was produced by electroplating gold conductors on both
sides of a $250\mu\text{m}$ thick aluminium oxide substrate of
size 15.7mm x 55.7mm. A tungsten-titanium alloy adhesive layer was
used between the substrate and the gold wires. The conductors are
$100\mu\text{m}$ wide and $6\mu\text{m}$ thick. Conductor
lengths vary from 2.3cm up to 6.5cm. In a vacuum environment, the
wires can sustain continuous currents of 1.35A,
corresponding to a current density of $j_\text{max} = 2.3\cdot
10^5 \text{A/cm}^2$. In pulsed operation (3s operation time, 60s
duty cycle), a maximum current of 2.8A and a current density of
$j_\text{max} = 4.7\cdot 10^5 \text{A/cm}^2$ was achieved.

The conductors QP1, QP2 and QP3 are each separated by a distance
of $d_\text{QP}=750\mu\text{m}$. Conductors T1-T8 underneath the
substrate are separated by $d_T=650\mu\text{m}$ and are repeated
periodically over a distance of 20.15mm. Crossings are avoided by
running these from the underneath to the top of the substrate and
back again. Contact is made through laser cut holes of
$400\mu\text{m}$ diameter and electroplating gold inside.
Electrical contacts to macroscopic conductors are made via the 24
contact pads on the top of the substrate.

A typical trap we use for the preparation and manipulation of
Bose-Einstein condensates of $^{87}\text{Rb}$ atoms in the F=2,
m$_{\text{F}}$=2 hyperfine ground state is generated by the
currents $I_{\text{QP1}}=I_{\text{QP3}}=0.85\text{A}$,
$I_{\text{QP2}}=0.235\text{A}$ and
$I_{\text{T1}}=I_{\text{T5}}=0.6\text{A}$. This trap is located
$300\mu\text{m}$ above the surface of the chip, the trap
frequencies are 140 Hz in the radial and 16 Hz in the axial
direction. The trap depth is 87$\mu K$ in the radial and
108$\mu K$ in the axial direction. In this trap, no fragmentation
of Bose-Einstein condensates is observed.

\subsection{Atom-optics Chips}
\label{sec:level4}

\begin{figure}
\centerline{\scalebox{0.65}{\includegraphics{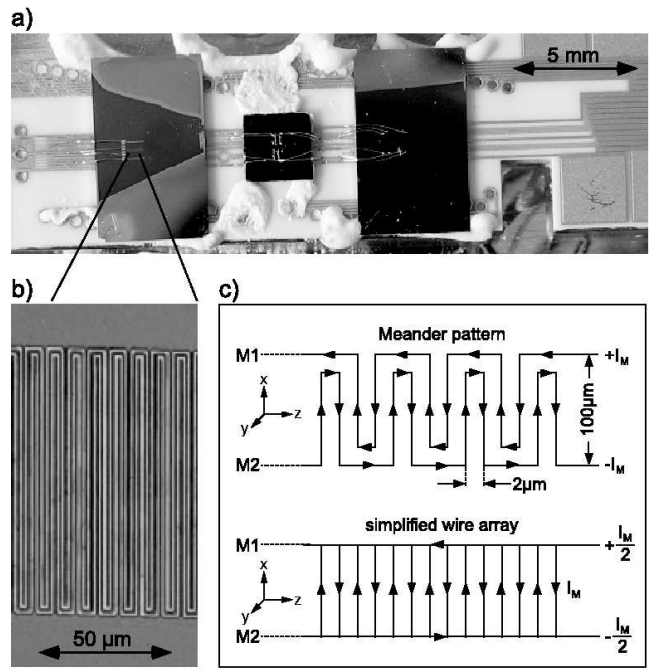}}}
\caption{(Color online) a) Combined chip. Atom-optics chips with micron scaled
conductor patterns are attached to the surface of the carrier
chip. b) Microscope image of a meandering conductor structure
deposited on the atom-optics chip on the left. Conductors of
$1\mu$m width are separated by gaps of $1\mu$m. c) Schematics and
connection scheme of the nested  meander patterns M1 and M2. The
simplified geometry (bottom) is used for calculating magnetic
fields in the center area of the structure.} \label{fig:Meander}
\end{figure}

Our atom-optics chips consist of micron-scaled conductors
deposited on silicon substrates. High quality 300nm thick gold
conductors have been patterned using standard techniques of
electron-beam lithography and dry etching. Figure
\ref{fig:Meander}a) and b) show the microscope image and
connection scheme of two nested meandering current patterns. The
$1\mu$m wide conductors are separated by $1\mu$m gaps. With 372
parallel conductor stripes, the structure has a total length of
$743\mu$m and a width of $100\mu$m. Driven with currents as
indicated in Figure \ref{fig:Meander}c, a magnetic lattice
potential \cite{Lau1999a,Lau1999b,Cognet1999a} of $4\mu$m period is generated.

Atom-optics chips are attached to the carrier using ceramic glue
(Figure \ref{fig:Meander}a). The meandering current pattern sits
approximately 250\,$\mu$m above the surface of the carrier chip.
The magnetic field of the meander can be superposed on the
magnetic waveguide potential positioned above it.

Because of its large size, the carrier chip can host several atom
optics chips, as depicted in Figure \ref{fig:Meander}a. The
electrical connection of atom-optics chips is realized by wire
bonding to the carrier chip. Atomic clouds initially loaded into
the magnetic potential of the carrier chip can be positioned over
any of the atom-optics chips using the conveyor and the waveguide.
Precise positioning of Bose-Einstein condensates on the
atom-optics chips is described in section IV.

\section{\label{sec:level6}Experiments on the carrier chip}

\subsection{\label{sec:level7}Transport on conveyor}

\begin{figure}
\centerline{\scalebox{0.9}{\includegraphics{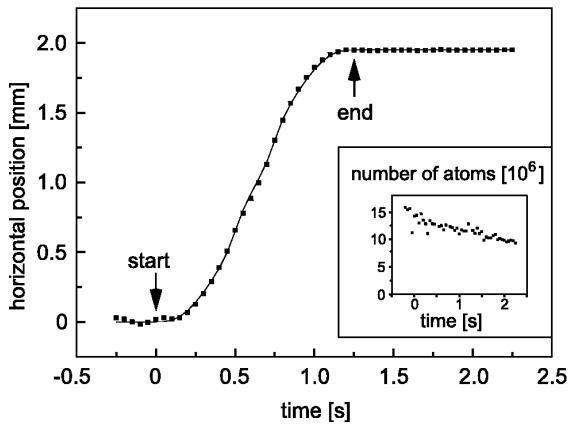}}}
\caption{Position of a thermal cloud during transport over a
distance of 1.95mm. The transport starts at $t=0$. The atoms are 
accelerated to a velocity of $v_T=2.6\text{mm/s}$
over 500ms. For the next 250ms the atoms move at constant velocity
$v_T$ before they are decelerated within another 500ms until they
come to rest at $t=1.25s$. The solid line shows the expected
position, calculated from the currents in the wires. The inset
image shows the number of atoms estimated from the absorption
images changing in time.} \label{fig:Transport}
\end{figure}

A key feature of the carrier chip is its ability to achieve smooth
transport over distances, limited only by the length of the chip,
in our experiment to 17.5mm.  Transport is demonstrated using a thermal 
cloud of $1.5\cdot 10^6$ rubidium atoms at a temperature of $6
\mu$K. The cloud is initially loaded into a trap as characterized
in the last paragraph of section III. A. We begin the transport by
accelerating the cloud over 500ms to a velocity of
$v_T=2.6\text{mm/s}$. The acceleration is increased to a maximum value of 
$a_\text{max}=8.2\text{mm/$s^2$}$ and subsequently reduced to zero, both with a sinusoidal characteristic shape. 
During the 500ms of
acceleration, the trap moves by $650\mu\text{m}$ along the
$z$-axis. Then the cloud moves with a constant velocity of
$v_T$=2.6\,mm/s and covers a distance of $650\,\mu\text{m}$ in
250\,ms. In the experiment shown in Figure \ref{fig:Transport},
the cloud was slowed down in 500\,ms using the inverse acceleration
ramp. The cloud arrives at rest (without sloshing) after
travelling a total distance of 1.95\,mm.

A thermal cloud is less sensitive to variations of the trap
parameters so this transport was done with a simplified, linear
relation between the currents $I_\text{T2}=I_\text{T}-I_\text{T1}$
instead of Eq. \ref{eq:CurrentForSmootTransport}. The displacement is then approximated by a
linear function of the currents in the transport wires, resulting
in a relevant simplification of the computer program controling
the experiment. However, this simplification means that the trap
parameters change during the transport: the radial trap frequency
decreases by up to 7 percent and the axial frequency increases by up to
14 percent. The dashed line in Figure \ref{fig:Potential}(left) shows
the axial potential shape for the simplified transport with
respect to the exact scheme, shown as a solid line. Nevertheless,
the transport takes place without any heating, since the atoms can
follow the frequency changes adiabatically. Absorption images of
the cloud taken every 50ms for a total transport time of 1.25s
were fitted by a two dimensional gaussian function in order to
determine the number of atoms, temperature and position. The five
absorption images in Figure \ref{fig:Potential} show the
acceleration period of the transport (first 500ms), while in
Figure \ref{fig:Transport} the measured position of the cloud
center (filled squares) as well as its numerically calculated
value (solid line) is plotted for the duration of the transport
(1.25\,s). The data and a calculation based on the current slopes
show excellent agreement. The inset of Figure \ref{fig:Transport}
shows the number of atoms in the trap over the transport period.
No relevant loss related to the transport is observed. A slow
evaporation and reduction of temperature is observed, which we
attribute to the reduced trap depth during transport.

In most cases, it is preferable to transport thermal clouds, and
subsequently generate condensates in the microtrap where further
experiments are to be performed. This way, no density dependent
losses or nonlinear excitations of the condensate are present
during the transport. Additional, micron-scaled positioning can be
done after producing the condensate. This is necessary when a
condensate has to be positioned a few microns from the surface, as
the radius of a thermal cloud exceeds this distance.

\subsection{\label{sec:level8}Position and momentum control of condensates}

\begin{figure}
\centerline{\scalebox{0.9}{\includegraphics{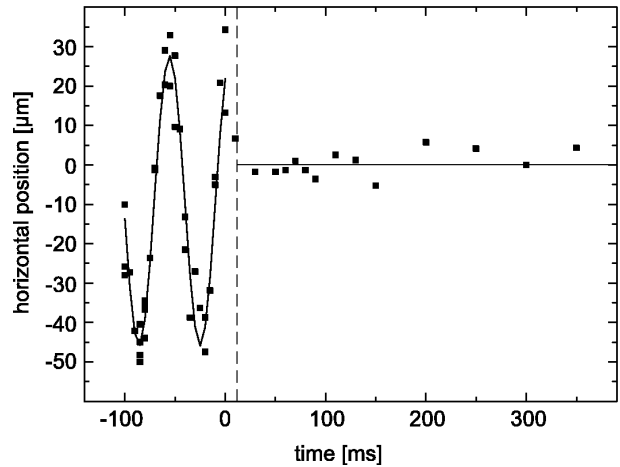}}}
\caption{Eliminating axial center-of-mass oscillations of a
Bose-Einstein condensate. Axial position of a condensate,
initially oscillating with $\omega=2\pi\cdot(16.4\pm 0.3)\text{ s}^{-1}$ and 
\mbox{$A=(36.8\pm 1.6)\,\mu\text{m}$}, after 25\,ms time-of-flight. The
center off mass motion is stopped by a phase-matched displacement
of the trap center using the conveyor.}
\label{fig:Oscillation}
\end{figure}
The carrier chip can be used for submicron positioning and
momentum control of the condensate. We demonstrate this by showing
that an initially sloshing condensate can be brought to rest by
carefully displacing the trap.  A small amplitude (harmonic)
oscillation is first induced on the trapped condensate
\begin{equation}
z(t)=A\sin\left(\omega_a t+\phi\right)+z_0\;.
\end{equation}
The phase and amplitude of the oscillation are detected after 25ms
time-of-flight (Figure \ref{fig:Oscillation}). As the imaged cloud position depends
on both its position and momentum at the time the trap is turned off, the data is fitted
by the function
\begin{eqnarray}
\begin{aligned}
Z(t)&=A\sin \left(\omega_a t+\phi\right)+A\omega_a\tau\cos\left(\omega_a t+\phi\right)+z_0\;,
\end{aligned}
\end{eqnarray}
describing the position of the oscillating cloud after $\tau$
time-of-flight. The fit parameters $A=(15.8\pm 0.8)\mu\text{m}$,
$\omega=2\pi\cdot(16.4\pm 0.3)\,\text{s}^{-1}$ and $\phi=(-0.014\pm 0.09)$ are used
to determine the magnitude and timing of the trap displacement
required to stop the oscillation. Radial excitations are avoided
if the displacement is applied on a time scale longer than the
radial oscillation period. The displacement is initiated 5\,ms
prior to when the center of mass motion reaches its turning point.
Within 10\,ms the trap is displaced by 16 microns along $z$, and
the cloud arrives at rest (Figure \ref{fig:Oscillation}). Based on the accuracy of the current
sources used for the experiment \cite{HighFinesse}, it is
in principle possible to calculate the trap position to a few nm
\cite{Hommelhoff2005a}. The accuracy of our measurement however,
is limited by the optical resolution of the imaging system which
is approximately $5\mu\text{m}$.

\section{\label{sec:level9}Experiments on atom-optics chips}
\subsection{\label{sec:level10}Positioning condensates on micron scaled atom-optics chips}

The time-of-flight position of a Bose-Einstein condensate is
determined by the motion of the cloud after the trapping potential
is turned off. Spin polarized atomic clouds in a magnetic field
sensitive state are accelerated not only by gravity but also by
magnetic field gradients. The method we describe for positioning
Bose-Einstein condensates on the micron scaled atom-optics chips
exploits the change in position of the condensate, after being
exposed to the magnetic field of the micro-structured
electromagnets, during the time of flight. The field adds to ambient
stray magnetic fields in the laboratory
and changes the trajectories of released clouds. In a series of
experiments, the initial position of the condensate can be varied
and the magnetic field of the atom-optics chip mapped out by
comparing the position after ballistic expansion with and without current in
the electromagnets.

\begin{figure}
\centerline{\scalebox{1.05}{\includegraphics{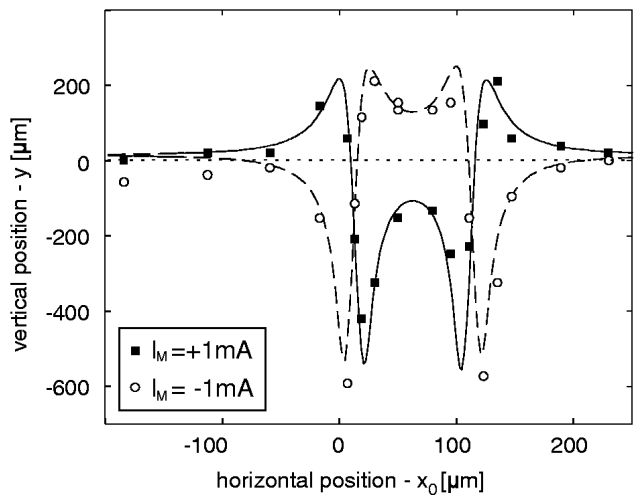}}}
\caption{Positioning condensates on a meandering wire pattern.
Plot of the position of Bose-Einstein condensates after 25 ms
ballistic expansion for currents $I_M=+1\text{mA}$ (solid squares) and
$I_M=-1\text{mA}$ (open circles) in the meandering current path,
relative to the position for zero current. The current was applied
during the first 5ms of the time-of-flight. The data is compared to
the numerical integration of the equation of motion,
based on the known structure of the
meander pattern (solid/dashed lines).} \label{fig:TrapCalibration}
\end{figure}

The method is demonstrated for an atom-optics chip with a known
meandering current pattern (Figure \ref{fig:Meander}). Using the
carrier chip, condensates were positioned at constant $y$- and
$z$-, but different $x$-positions below the chip. At each
horizontal position, the distance to the surface was calibrated by
moving the condensate towards the chip surface, until it was lost
due to atom-surface interactions. During positioning, the
meandering current was turned off. For each initial $x$-position,
the vertical position ($y$) of the condensate was detected after
25\,ms time-of-flight with zero and $\pm 1\text{mA}$ current in
the meander. The current was applied during the first 5\,ms of the
ballistic expansion. Because the imaging beam propagates parallel
to the $x$-axis, only displacements along $y$ and $z$ were
detectable. Measurements in the center area of the meander reveal
only vertical $y$-displacements. The displacements plotted in
Figure \ref{fig:TrapCalibration} are relative to the
time-of-flight position for zero current. Although the cloud was
only shifted parallel to the imaging beam, we were able to
accurately detect the $100\,\mu$m wide feature of the meander. 
Due to the finite width of the pattern ($100\,\mu\text{m}$), 
the characteristics are dominated by the magnetic field of currents 
at the edge of the meander (see Figure \ref{fig:Meander}, simplified wire array) 
while the contribution of the periodic pattern with alternating currents 
is negligible for distances larger than the period \cite{Lau1999b}. 
The asymmetry for positive and negative currents is due to ambient magnetic fields. 

The data is compared to the numerical integration of the equation of motion for 
rubidium atoms in this magnetic field. The simultaneous fit to both data sets, for 1\,mA and -1\,mA, gives
the center position of the meander structure to be shifted
by $x_0=62.58 \pm 1.37 \mu\text{m}$ relative to the origin which was
defined to be the center of the conductor QP2 on the carrier chip.
The fit also gives the distance to the chip surface $y_0=14.00 \pm
1.55 \mu\text{m}$ and the ambient magnetic fields ($B_x\approx 0 G, B_y\approx 1 G, B_z\approx 1.9 G$).

The same method is applicable for measuring the position of the
meandering current pattern along the $z$-coordinate. Calibration
curves such as in Figure \ref{fig:TrapCalibration} allow the
ultra-precise, three-dimensional positioning of condensates on
micron scaled atom-optics chips.

\subsection{\label{sec:level11}Direction sensitive magnetic field microscope}

\begin{figure}
\centerline{\scalebox{0.9}{\includegraphics{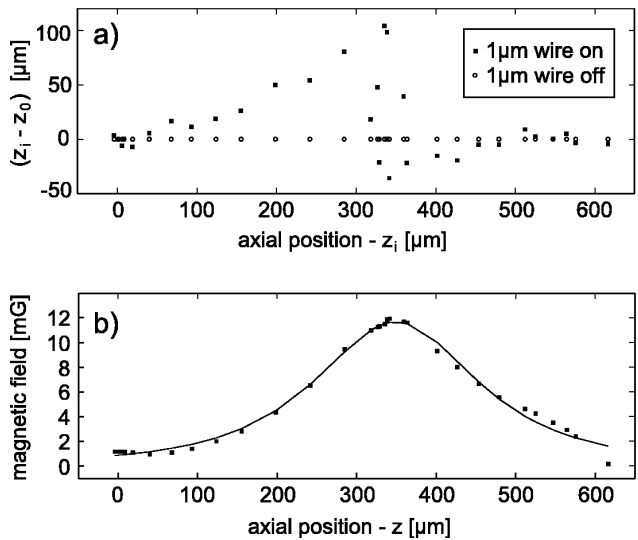}}}
\caption{a) Magnetic field microscope. Positions $z_0$ (open
circles) and $z_i$ (filled squares) of the Bose-Einstein
condensate for zero and for 1.2mA current, respectively, in a thin
conductor perpendicular to the waveguide. b)  $z$-component of the
magnetic field of the thin conductor, as integrated from the
measured gradient according Eq. \ref{eq:microscopegradient}. The data is in good agreement
with the axial field component calculated for a finite wire
oriented parallel to the $x$-axis.} \label{fig:Mikroskop}
\end{figure}

We now describe the principle of a magnetic field microscope based
on the controlled translation of a well known magnetic potential.
Bose-Einstein condensates are used as tiny probes,
for measuring the difference between the applied and the actual
potential. The microscope is demonstrated by mapping the magnetic
field of a one micron wide test conductor.

The potential experienced by the condensate can be decomposed into
a well known trapping potential $U_0(z)$ and an unknown potential
$U_1 (z)$ to be measured:
\begin{equation}
U(z)=U_0(z)+U_1(z)\;.
\end{equation}
The potential $U_0(z)$ is related to the axial confinement of our
waveguide on the carrier chip
\begin{equation}
U_0(z)=\frac{1}{2}m\omega^2(z-z_0)^2\;.
\end{equation}
At the minimum $z_i$ of the potential U(z), the first derivative of $U
(z)$ vanishes and
\begin{equation}
\label{eq:microscopegradient}
\left.\frac{dU_1(z)}{dz}\right|_{z_i}=-\left.\frac{dU_0(z)}{dz}\right|_{z_i}=-m\omega^2(z_i-z_0)
\end{equation}
with $z_0$ being the minimum of $U_0$.

According to Eq. \ref{eq:microscopegradient}, the gradient of $U_1(z_i)$ is proportional to
the difference $(z_i-z_0)$. From this difference, the magnetic field
distribution of the sample can be derived. By translating the center $z_0$ of the
well known potential $U_0(z)$ and detecting the position $z_i$
where a Bose-Einstein condensate is localized, the gradient of
$U_1$ can be measured as a function of $z_i$. Note, that the
potential $U(z)$ can have several minima $z_i$, ($i=1,2,...$).
Integration of Eq. \ref{eq:microscopegradient} over $z$ gives the potential $U_1(z)$.

The realization of this microscope requires the potential $U_0(z)$
to be well known. It must also be possible to precisely shift its position
which is possible with the micro structured conveyor belt of the carrier chip
described in section II. Keeping the $x$- and $y$-coordinates
constant is essential for the measurement. Keeping also the axial
oscillation frequency $\omega_a$ constant simplifies the
analysis. The conveyor belt and manipulation schemes described in
this article easily meet these requirements. Through the
three-dimensional position control of Bose-Einstein condensates on
the carrier chip, a three-dimensional, highly sensitive
measurement of the independent magnetic field components is
possible. This in turn allows the complete mapping of unknown
magnetic potentials.

We demonstrate the microscope by measuring the $z$-component of a
$1\mu\text{m}$ wide, $300\mu\text{m}$ long micro structured
conductor aligned parallel to the $x$-axis. The conductor was
patterned on the surface of an atom-optics chip, the same as used
for the meandering pattern. Experiments were carried out
$390\,\mu\text{m}$ away from the carrier chip in a smooth
waveguide potential with an axial oscillation frequency of 16 Hz.
Bose-Einstein condensates were positioned at different
$z$-positions around the thin conductor, and the position $z_i$
was detected for a current of 1.2mA, and $z_0$ for a current zero,
as shown in Figure \ref{fig:Mikroskop}a. The position difference
taken as a function of $z_i$ was integrated to find the magnetic
field curve \ref{fig:Mikroskop}b of the single conductor.  The
solid line is the calculated field of a finite length wire located
$140\mu\text{m}$ below the condensate. The sensitivity of the microscope 
can be calculated with Eq. \ref{eq:microscopegradient} and depends only 
on the trap frequency and the resolution of the imaging system. With a reasonably 
low trap frequency of 1Hz and an optical resolution of $1\mu\text{m}$ a sensitivity 
of about $6.2\cdot 10^{-5}\text{G/cm}$ should be possible. In our experiment with $\omega_a = 2\pi\cdot 16\text{s}^{-1}$, 
and an optical resolution of $5\,\mu\text{m}$, the sensitivity was 0.08 G/cm.

\section{\label{sec:level12}Conclusions}

We have presented experiments using a combined atom chip which
allows one to work a few microns from a surface.  Fragmentation is
avoided because the waveguide potential is formed by conductors at
large distances from the condensate.  Johnson-noise induced spin
flip losses are greatly reduced as the only conductor in close
proximity is a narrow wire aligned transverse to the cloud
\cite{Henkel2001a}.
The described methods here provide a route towards micron scale
atom optics.

We have also performed experiments in which condensates were
loaded into a magnetic lattice produced by the meandering current
pattern. While interference following release from these traps was
observed \cite{Guenther2005b}, the (exponential) height dependence
of the lattice leads to a highly complex situation.  This
contrasts with optical standing waves where the lattice potential
is homogeneous over the size of the condensate. It therefore
remains an experimental challenge to engineer advanced
atom-optical elements with magnetic potentials for extended clouds
of Bose-Einstein condensates.  Many of the proposed advantages of
atom chips, however, such as precisely controlled tunnelling rates
through potential barriers and in lattices should be attainable
using smaller wave packets or single atoms.

\begin{acknowledgments}
This work was supported by the Deutsche Forschungsgemeinschaft,
Landesstiftung Baden-W\"urttemberg, EU Marie-Curie RTN on Atomchips,
and the Australian Research Council. The authors thank D. K\"olle, R. Kleiner (Physikalisches Institut, Universit\"at T\"ubingen), 
and C. D\"ucs\H{o} (MFA/KFKI) for production of micro chips.
\end{acknowledgments}


\begin{thebibliography}{23}
\expandafter\ifx\csname natexlab\endcsname\relax\def\natexlab#1{#1}\fi
\expandafter\ifx\csname bibnamefont\endcsname\relax
  \def\bibnamefont#1{#1}\fi
\expandafter\ifx\csname bibfnamefont\endcsname\relax
  \def\bibfnamefont#1{#1}\fi
\expandafter\ifx\csname citenamefont\endcsname\relax
  \def\citenamefont#1{#1}\fi
\expandafter\ifx\csname url\endcsname\relax
  \def\url#1{\texttt{#1}}\fi
\expandafter\ifx\csname urlprefix\endcsname\relax\def\urlprefix{URL }\fi
\providecommand{\bibinfo}[2]{#2}
\providecommand{\eprint}[2][]{\url{#2}}

\bibitem[{\citenamefont{Weinstein and Libbrecht}(1995)}]{Weinstein1995a}
\bibinfo{author}{\bibfnamefont{J.~D.} \bibnamefont{Weinstein}}
  \bibnamefont{and} \bibinfo{author}{\bibfnamefont{K.~G.}
  \bibnamefont{Libbrecht}}, \bibinfo{journal}{Phys. Rev. A}
  \textbf{\bibinfo{volume}{52}}, \bibinfo{pages}{4004} (\bibinfo{year}{1995}).

\bibitem[{\citenamefont{Smerzi et~al.}(1997)\citenamefont{Smerzi, Fantoni,
  Giovanazzi, and Shenoy}}]{Smerzi1997a}
\bibinfo{author}{\bibfnamefont{A.}~\bibnamefont{Smerzi}},
  \bibinfo{author}{\bibfnamefont{S.}~\bibnamefont{Fantoni}},
  \bibinfo{author}{\bibfnamefont{S.}~\bibnamefont{Giovanazzi}},
  \bibnamefont{and} \bibinfo{author}{\bibfnamefont{S.~R.}
  \bibnamefont{Shenoy}}, \bibinfo{journal}{Phys. Rev. Lett.}
  \textbf{\bibinfo{volume}{79}}, \bibinfo{pages}{4950} (\bibinfo{year}{1997}).

\bibitem[{\citenamefont{Paul et~al.}(2005)\citenamefont{Paul, Richter, and
  Schlagheck}}]{Paul2005a}
\bibinfo{author}{\bibfnamefont{T.}~\bibnamefont{Paul}},
  \bibinfo{author}{\bibfnamefont{K.}~\bibnamefont{Richter}}, \bibnamefont{and}
  \bibinfo{author}{\bibfnamefont{P.}~\bibnamefont{Schlagheck}},
  \bibinfo{journal}{Phys. Rev. Lett.} \textbf{\bibinfo{volume}{94}},
  \bibinfo{pages}{020404} (\bibinfo{year}{2005}).

\bibitem[{\citenamefont{Dorner et~al.}(2003)\citenamefont{Dorner, Fedichev,
  Jaksch, Lewenstein, and Zoller}}]{Dorner2003a}
\bibinfo{author}{\bibfnamefont{U.}~\bibnamefont{Dorner}},
  \bibinfo{author}{\bibfnamefont{P.}~\bibnamefont{Fedichev}},
  \bibinfo{author}{\bibfnamefont{D.}~\bibnamefont{Jaksch}},
  \bibinfo{author}{\bibfnamefont{M.}~\bibnamefont{Lewenstein}},
  \bibnamefont{and} \bibinfo{author}{\bibfnamefont{P.}~\bibnamefont{Zoller}},
  \bibinfo{journal}{Phys. Rev. Lett.} \textbf{\bibinfo{volume}{91}},
  \bibinfo{pages}{073601} (\bibinfo{year}{2003}).

\bibitem[{\citenamefont{Ott et~al.}(2001)\citenamefont{Ott, Fort\'{a}gh,
  Schlotterbeck, Grossmann, and Zimmermann}}]{Ott2001a}
\bibinfo{author}{\bibfnamefont{H.}~\bibnamefont{Ott}},
  \bibinfo{author}{\bibfnamefont{J.}~\bibnamefont{Fort\'{a}gh}},
  \bibinfo{author}{\bibfnamefont{G.}~\bibnamefont{Schlotterbeck}},
  \bibinfo{author}{\bibfnamefont{A.}~\bibnamefont{Grossmann}},
  \bibnamefont{and}
  \bibinfo{author}{\bibfnamefont{C.}~\bibnamefont{Zimmermann}},
  \bibinfo{journal}{Phys. Rev. Lett.} \textbf{\bibinfo{volume}{87}},
  \bibinfo{pages}{230401} (\bibinfo{year}{2001}).

\bibitem[{\citenamefont{H{\"a}nsel et~al.}(2001)\citenamefont{H{\"a}nsel,
  Hommelhoff, H{\"a}nsch, and Reichel}}]{Haensel2001a}
\bibinfo{author}{\bibfnamefont{W.}~\bibnamefont{H{\"a}nsel}},
  \bibinfo{author}{\bibfnamefont{P.}~\bibnamefont{Hommelhoff}},
  \bibinfo{author}{\bibfnamefont{T.~W.} \bibnamefont{H{\"a}nsch}},
  \bibnamefont{and} \bibinfo{author}{\bibfnamefont{J.}~\bibnamefont{Reichel}},
  \bibinfo{journal}{Nature} \textbf{\bibinfo{volume}{413}},
  \bibinfo{pages}{498} (\bibinfo{year}{2001}).

\bibitem[{\citenamefont{Henkel et~al.}(1999)\citenamefont{Henkel, P{\"o}tting,
  and Wilkens}}]{Henkel1999a}
\bibinfo{author}{\bibfnamefont{C.}~\bibnamefont{Henkel}},
  \bibinfo{author}{\bibfnamefont{S.}~\bibnamefont{P{\"o}tting}},
  \bibnamefont{and} \bibinfo{author}{\bibfnamefont{M.}~\bibnamefont{Wilkens}},
  \bibinfo{journal}{Appl. Phys. B} \textbf{\bibinfo{volume}{69}},
  \bibinfo{pages}{379} (\bibinfo{year}{1999}).

\bibitem[{\citenamefont{Jones et~al.}(2004)\citenamefont{Jones, Vale, Sahagun,
  Hall, Eberlein, Sauer, Furusawa, Richardson, and Hinds}}]{Jones2004a}
\bibinfo{author}{\bibfnamefont{M.~P.~A.} \bibnamefont{Jones}},
  \bibinfo{author}{\bibfnamefont{C.~J.} \bibnamefont{Vale}},
  \bibinfo{author}{\bibfnamefont{D.}~\bibnamefont{Sahagun}},
  \bibinfo{author}{\bibfnamefont{B.~V.} \bibnamefont{Hall}},
  \bibinfo{author}{\bibfnamefont{C.~C.} \bibnamefont{Eberlein}},
  \bibinfo{author}{\bibfnamefont{B.~E.} \bibnamefont{Sauer}},
  \bibinfo{author}{\bibfnamefont{K.}~\bibnamefont{Furusawa}},
  \bibinfo{author}{\bibfnamefont{D.}~\bibnamefont{Richardson}},
  \bibnamefont{and} \bibinfo{author}{\bibfnamefont{E.~A.} \bibnamefont{Hinds}},
  \bibinfo{journal}{J. Phys. B} \textbf{\bibinfo{volume}{37}},
  \bibinfo{pages}{L15} (\bibinfo{year}{2004}).

\bibitem[{\citenamefont{Est\`eve et~al.}(2004)\citenamefont{Est\`eve, Aussibal,
  Schumm, Figl, Mailly, Bouchoule, Westbrook, and Aspect}}]{Estve2004a}
\bibinfo{author}{\bibfnamefont{J.}~\bibnamefont{Est\`eve}},
  \bibinfo{author}{\bibfnamefont{C.}~\bibnamefont{Aussibal}},
  \bibinfo{author}{\bibfnamefont{T.}~\bibnamefont{Schumm}},
  \bibinfo{author}{\bibfnamefont{C.}~\bibnamefont{Figl}},
  \bibinfo{author}{\bibfnamefont{D.}~\bibnamefont{Mailly}},
  \bibinfo{author}{\bibfnamefont{I.}~\bibnamefont{Bouchoule}},
  \bibinfo{author}{\bibfnamefont{C.~I.} \bibnamefont{Westbrook}},
  \bibnamefont{and} \bibinfo{author}{\bibfnamefont{A.}~\bibnamefont{Aspect}},
  \bibinfo{journal}{Phys. Rev. A} \textbf{\bibinfo{volume}{70}},
  \bibinfo{pages}{043629} (\bibinfo{year}{2004}).

\bibitem[{\citenamefont{Kraft et~al.}(2002)\citenamefont{Kraft, G{\"u}nther,
  Ott, Wharam, Zimmermann, and Fort\'agh}}]{Kraft2002a}
\bibinfo{author}{\bibfnamefont{S.}~\bibnamefont{Kraft}},
  \bibinfo{author}{\bibfnamefont{A.}~\bibnamefont{G{\"u}nther}},
  \bibinfo{author}{\bibfnamefont{H.}~\bibnamefont{Ott}},
  \bibinfo{author}{\bibfnamefont{D.}~\bibnamefont{Wharam}},
  \bibinfo{author}{\bibfnamefont{C.}~\bibnamefont{Zimmermann}},
  \bibnamefont{and}
  \bibinfo{author}{\bibfnamefont{J.}~\bibnamefont{Fort\'agh}},
  \bibinfo{journal}{J. Phys. B} \textbf{\bibinfo{volume}{35}},
  \bibinfo{pages}{L469} (\bibinfo{year}{2002}).

\bibitem[{\citenamefont{Ott et~al.}(2003)\citenamefont{Ott, Fort\'agh, Kraft,
  G{\"u}nther, Komma, and Zimmermann}}]{Ott2003a}
\bibinfo{author}{\bibfnamefont{H.}~\bibnamefont{Ott}},
  \bibinfo{author}{\bibfnamefont{J.}~\bibnamefont{Fort\'agh}},
  \bibinfo{author}{\bibfnamefont{S.}~\bibnamefont{Kraft}},
  \bibinfo{author}{\bibfnamefont{A.}~\bibnamefont{G{\"u}nther}},
  \bibinfo{author}{\bibfnamefont{D.}~\bibnamefont{Komma}}, \bibnamefont{and}
  \bibinfo{author}{\bibfnamefont{C.}~\bibnamefont{Zimmermann}},
  \bibinfo{journal}{Phys. Rev. Lett.} \textbf{\bibinfo{volume}{91}},
  \bibinfo{pages}{040402} (\bibinfo{year}{2003}).

\bibitem[{\citenamefont{Henkel et~al.}(2003)\citenamefont{Henkel, Kr{\"u}ger,
  Folman, and Schmiedmayer}}]{Henkel2003}
\bibinfo{author}{\bibfnamefont{C.}~\bibnamefont{Henkel}},
  \bibinfo{author}{\bibfnamefont{P.}~\bibnamefont{Kr{\"u}ger}},
  \bibinfo{author}{\bibfnamefont{R.}~\bibnamefont{Folman}}, \bibnamefont{and}
  \bibinfo{author}{\bibfnamefont{J.}~\bibnamefont{Schmiedmayer}},
  \bibinfo{journal}{Appl. Phys. B} \textbf{\bibinfo{volume}{76}},
  \bibinfo{pages}{173} (\bibinfo{year}{2003}).

\bibitem[{\citenamefont{Folman et~al.}(2002)\citenamefont{Folman, Kr{\"u}ger,
  Schmiedmayer, Denschlag, and Henkel}}]{Folman2002a}
\bibinfo{author}{\bibfnamefont{R.}~\bibnamefont{Folman}},
  \bibinfo{author}{\bibfnamefont{P.}~\bibnamefont{Kr{\"u}ger}},
  \bibinfo{author}{\bibfnamefont{J.}~\bibnamefont{Schmiedmayer}},
  \bibinfo{author}{\bibfnamefont{J.}~\bibnamefont{Denschlag}},
  \bibnamefont{and} \bibinfo{author}{\bibfnamefont{C.}~\bibnamefont{Henkel}},
  \bibinfo{journal}{Adv. At. Mol. Opt. Phys.} \textbf{\bibinfo{volume}{48}},
  \bibinfo{pages}{263} (\bibinfo{year}{2002}).

\bibitem[{\citenamefont{Sukumar and Brink}(1997)}]{Sukumar1997a}
\bibinfo{author}{\bibfnamefont{C.~V.} \bibnamefont{Sukumar}} \bibnamefont{and}
  \bibinfo{author}{\bibfnamefont{D.~M.} \bibnamefont{Brink}},
  \bibinfo{journal}{Phys. Rev. A} \textbf{\bibinfo{volume}{56}},
  \bibinfo{pages}{2451} (\bibinfo{year}{1997}).

\bibitem[{\citenamefont{Pritchard}(1983)}]{Pritchard1983a}
\bibinfo{author}{\bibfnamefont{D.~E.} \bibnamefont{Pritchard}},
  \bibinfo{journal}{Phys. Rev. Lett.} \textbf{\bibinfo{volume}{51}},
  \bibinfo{pages}{1336} (\bibinfo{year}{1983}).

\bibitem[{\citenamefont{Kraft et~al.}(2005)\citenamefont{Kraft, G{\"u}nther,
  Wicke, Kasch, Zimmermann, and Fort\'agh}}]{Guenther2005a}
\bibinfo{author}{\bibfnamefont{S.}~\bibnamefont{Kraft}},
  \bibinfo{author}{\bibfnamefont{A.}~\bibnamefont{G{\"u}nther}},
  \bibinfo{author}{\bibfnamefont{P.}~\bibnamefont{Wicke}},
  \bibinfo{author}{\bibfnamefont{B.}~\bibnamefont{Kasch}},
  \bibinfo{author}{\bibfnamefont{C.}~\bibnamefont{Zimmermann}},
  \bibnamefont{and}
  \bibinfo{author}{\bibfnamefont{J.}~\bibnamefont{Fort\'agh}},
  \bibinfo{journal}{cond-mat/0504242}  (\bibinfo{year}{unpublished}).

\bibitem[{\citenamefont{Lau et~al.}(1999{\natexlab{a}})\citenamefont{Lau,
  Sidorov, Opat, McLean, Rowlands, and Hannaford}}]{Lau1999a}
\bibinfo{author}{\bibfnamefont{D.~C.} \bibnamefont{Lau}},
  \bibinfo{author}{\bibfnamefont{A.~I.} \bibnamefont{Sidorov}},
  \bibinfo{author}{\bibfnamefont{G.~I.} \bibnamefont{Opat}},
  \bibinfo{author}{\bibfnamefont{R.~J.} \bibnamefont{McLean}},
  \bibinfo{author}{\bibfnamefont{W.~J.} \bibnamefont{Rowlands}},
  \bibnamefont{and}
  \bibinfo{author}{\bibfnamefont{P.}~\bibnamefont{Hannaford}},
  \bibinfo{journal}{Eur. Phys. J. D} \textbf{\bibinfo{volume}{5}},
  \bibinfo{pages}{193} (\bibinfo{year}{1999}{\natexlab{a}}).

\bibitem[{\citenamefont{Lau et~al.}(1999{\natexlab{b}})\citenamefont{Lau,
  McLean, Sidorov, Gough, Koperski, Rowlands, Sexton, Opat, and
  Hannaford}}]{Lau1999b}
\bibinfo{author}{\bibfnamefont{D.~C.} \bibnamefont{Lau}},
  \bibinfo{author}{\bibfnamefont{R.~J.} \bibnamefont{McLean}},
  \bibinfo{author}{\bibfnamefont{A.~I.} \bibnamefont{Sidorov}},
  \bibinfo{author}{\bibfnamefont{D.~S.} \bibnamefont{Gough}},
  \bibinfo{author}{\bibfnamefont{J.}~\bibnamefont{Koperski}},
  \bibinfo{author}{\bibfnamefont{W.~J.} \bibnamefont{Rowlands}},
  \bibinfo{author}{\bibfnamefont{B.~A.} \bibnamefont{Sexton}},
  \bibinfo{author}{\bibfnamefont{G.~I.} \bibnamefont{Opat}}, \bibnamefont{and}
  \bibinfo{author}{\bibfnamefont{P.}~\bibnamefont{Hannaford}},
  \bibinfo{journal}{J. Opt. B} \textbf{\bibinfo{volume}{1}},
  \bibinfo{pages}{371} (\bibinfo{year}{1999}{\natexlab{b}}).

\bibitem[{\citenamefont{{Cognet et al.}}(1999)}]{Cognet1999a}
\bibinfo{author}{\bibfnamefont{L.}~\bibnamefont{{Cognet et al.}}},
  \bibinfo{journal}{Europhys. Lett.} \textbf{\bibinfo{volume}{47}},
  \bibinfo{pages}{538} (\bibinfo{year}{1999}).

\bibitem[{Hig()}]{HighFinesse}
\urlprefix\url{http://www.highfinesse.de/steuer_1.html}.

\bibitem[{\citenamefont{Hommelhoff et~al.}(2005)\citenamefont{Hommelhoff,
  H{\"a}nsel, Steinmetz, H{\"a}nsch, and Reichel}}]{Hommelhoff2005a}
\bibinfo{author}{\bibfnamefont{P.}~\bibnamefont{Hommelhoff}},
  \bibinfo{author}{\bibfnamefont{W.}~\bibnamefont{H{\"a}nsel}},
  \bibinfo{author}{\bibfnamefont{T.}~\bibnamefont{Steinmetz}},
  \bibinfo{author}{\bibfnamefont{T.~W.} \bibnamefont{H{\"a}nsch}},
  \bibnamefont{and} \bibinfo{author}{\bibfnamefont{J.}~\bibnamefont{Reichel}},
  \bibinfo{journal}{New J. Phys.} \textbf{\bibinfo{volume}{7}},
  \bibinfo{pages}{3} (\bibinfo{year}{2005}).

\bibitem[{\citenamefont{Henkel and P{\"o}tting}(2001)}]{Henkel2001a}
\bibinfo{author}{\bibfnamefont{C.}~\bibnamefont{Henkel}} \bibnamefont{and}
  \bibinfo{author}{\bibfnamefont{S.}~\bibnamefont{P{\"o}tting}},
  \bibinfo{journal}{Appl. Phys. B} \textbf{\bibinfo{volume}{72}},
  \bibinfo{pages}{73} (\bibinfo{year}{2001}).

\bibitem[{\citenamefont{G{\"u}nther et~al.}(2005)\citenamefont{G{\"u}nther,
  Kraft, Kemmler, K{\"o}lle, Kleiner, Zimmermann, and
  Fort\'agh}}]{Guenther2005b}
\bibinfo{author}{\bibfnamefont{A.}~\bibnamefont{G{\"u}nther}},
  \bibinfo{author}{\bibfnamefont{S.}~\bibnamefont{Kraft}},
  \bibinfo{author}{\bibfnamefont{M.}~\bibnamefont{Kemmler}},
  \bibinfo{author}{\bibfnamefont{D.}~\bibnamefont{K{\"o}lle}},
  \bibinfo{author}{\bibfnamefont{R.}~\bibnamefont{Kleiner}},
  \bibinfo{author}{\bibfnamefont{C.}~\bibnamefont{Zimmermann}},
  \bibnamefont{and}
  \bibinfo{author}{\bibfnamefont{J.}~\bibnamefont{Fort\'agh}},
  \bibinfo{journal}{cond-mat/0504210}  (\bibinfo{year}{unpublished}).

\end{thebibliography}
\end{document}